\documentclass{PoS}

\title{Testing the Field Correlator Method with astrophysical constraints}

\ShortTitle{Testing the Field Correlator Method with astrophysical constraints}

\author{\speaker{{G. F. Burgio}}\\
        INFN Sezione di Catania, Via S. Sofia 64, I-95123 Catania, Italy\\
        E-mail: \email{fiorella.burgio@ct.infn.it}}

\author{M. Baldo\\
        INFN Sezione di Catania, Via S. Sofia 64, I-95123 Catania, Italy\\
        E-mail: \email{marcello.baldo@ct.infn.it}}

\author{P. Castorina\\
        Dipartimento di Fisica e Astronomia, Universita' di Catania, and 
        INFN Sezione di Catania, Via S. Sofia 64, I-95123 Catania, Italy\\
        E-mail: \email{paolo.castorina@ct.infn.it}}

\author{S. Plumari\\
        Dipartimento di Fisica e Astronomia, Universita' di Catania, and
        INFN Sezione di Catania, Via S. Sofia 64, I-95123 Catania, Italy\\
        E-mail: \email{salvatore.plumari@ct.infn.it}}

\author{D. Zappala'\\
        INFN Sezione di Catania, Via S. Sofia 64, I-95123 Catania, Italy\\
        E-mail: \email{dario.zappala@ct.infn.it}}

\abstract{We study the structure of hybrid stars with the Field Correlator
Method, extended to the zero temperature limit, for the quark phase.
For the hadronic phase, we use the microscopic Brueckner-Hartree-
Fock many-body theory. The comparison with the  neutron star 
mass phenomenology puts serious constraints on the currently adopted values 
of the gluon 
condensate $G_2 \simeq 0.006-0.007~\rm {GeV^4}$, and the large distance 
static $Q \bar Q$ potential.}

\FullConference{8th Conference Quark Confinement and the Hadron Spectrum \\
		 September 1-6 2008\\
		 Mainz, Germany}

\begin{document}

\section{Introduction}

The study of the properties of neutron stars (NS) concerns the large density
(and low temperature) region of the  phase diagram and, in particular,
it requires the QCD non-pertubative Equation of State (EoS) at small $T$ and
large $\mu_q$, where no QCD lattice simulations are available yet.
Due to the lack of lattice data, analytic approaches based on more elementary
models, such as the Nambu--Jona-Lasinio (NJL) model \cite{bubtes}, 
the MIT Bag model, and the Color Dielectric model \cite{col} 
are mostly used. A serious drawback of those models is that they cannot 
make predictions for both limits, i.e.high temperature and zero chemical
potential or high chemical potential and low temperature, and therefore
cannot be fully tested.
One of the few exceptions is the Field Correlator Method (FCM) \cite{phrep},
which in principle is able to cover the full temperature-chemical potential 
plane. The method contains {\it ab initio} the property of
confinement, which should play a role in the stability of the pure quark phase
in neutron stars, as we discussed in ref.\cite{noi1}.
Based on that, we have tested \cite{noi2} the FCM by comparing the results for 
the neutron star masses with the existing phenomenology, which turns out to
be a strong constraint on the parameters used in the model.
In particular, we have found definite numerical indications on some relevant
physical quantities, such as the gluon condensate and the $Q \bar Q$ potential, 
to be compared to the ones extracted from the determination of the
critical temperature of the deconfinement phase
transition. This shows the relevance of the comparison of the model predictions
in the high chemical potential region with the astrophysical phenomenology.

\section{Hadronic and quark EOS}
\par\noindent
The main point of our study is the comparison of the nuclear matter EoS with 
the one for quark matter.  
We  start  with the description of the hadronic phase. 
The EoS is based on the Brueckner--Bethe--Goldstone (BBG) many-body theory, 
which is a linked cluster expansion of the energy per nucleon of nuclear 
matter\cite{tri}. It has been shown
that the non-relativistic BBG expansion is well convergent,
and  the Brueckner-Hartree-Fock (BHF) level of approximation is accurate
in the density range relevant for neutron  stars.  
In the BHF approximation the energy per nucleon is
\begin{equation}
{E \over{A}}  =  
          {{3}\over{5}}{{k_F^2}\over {2m}}  + {{1}\over{2n}}  
~ {\rm Re} \sum_{k,k'\leq k_F} \langle k k'|G[n; e(k)+e(k')]|k k'\rangle_a. 
\end{equation}
\noindent
where $G$ is the Brueckner reaction matrix, which contains the bare 
nucleon-nucleon (NN) interaction, and the nucleon number density $n$.
$e(k)$ is the single-particle energy, and the subscript ``{\it a}'' indicates 
antisymmetrization of the matrix element.  
In the calculations reported here we have used the Argonne $v_{18}$ potential
as the two-nucleon interaction, supplemented by 
three-body forces built according to the Urbana model \cite{uix}.  
The corresponding nuclear matter EOS reproduces correctly the nuclear matter 
saturation point $\rho_0=0.17~\rm fm^{-3}$, $E/A=-16~\rm MeV$ \cite{bbb}.
In neutron stars one has to consider matter in beta equilibrium,
where electrons and eventually muons coexist with baryons,
while neutrinos are considered to escape from the star.
The EOS for the beta equilibrated matter can be obtained 
once the hadron matter is known, together with the chemical
potentials of different species as a function of total baryon 
density. Since the procedure is standard, we do not give further details 
of the calculations.
\par
\begin{figure}
\begin{picture}(30,200)
\put(50,200){
  \includegraphics[height=5.2cm,angle=270]{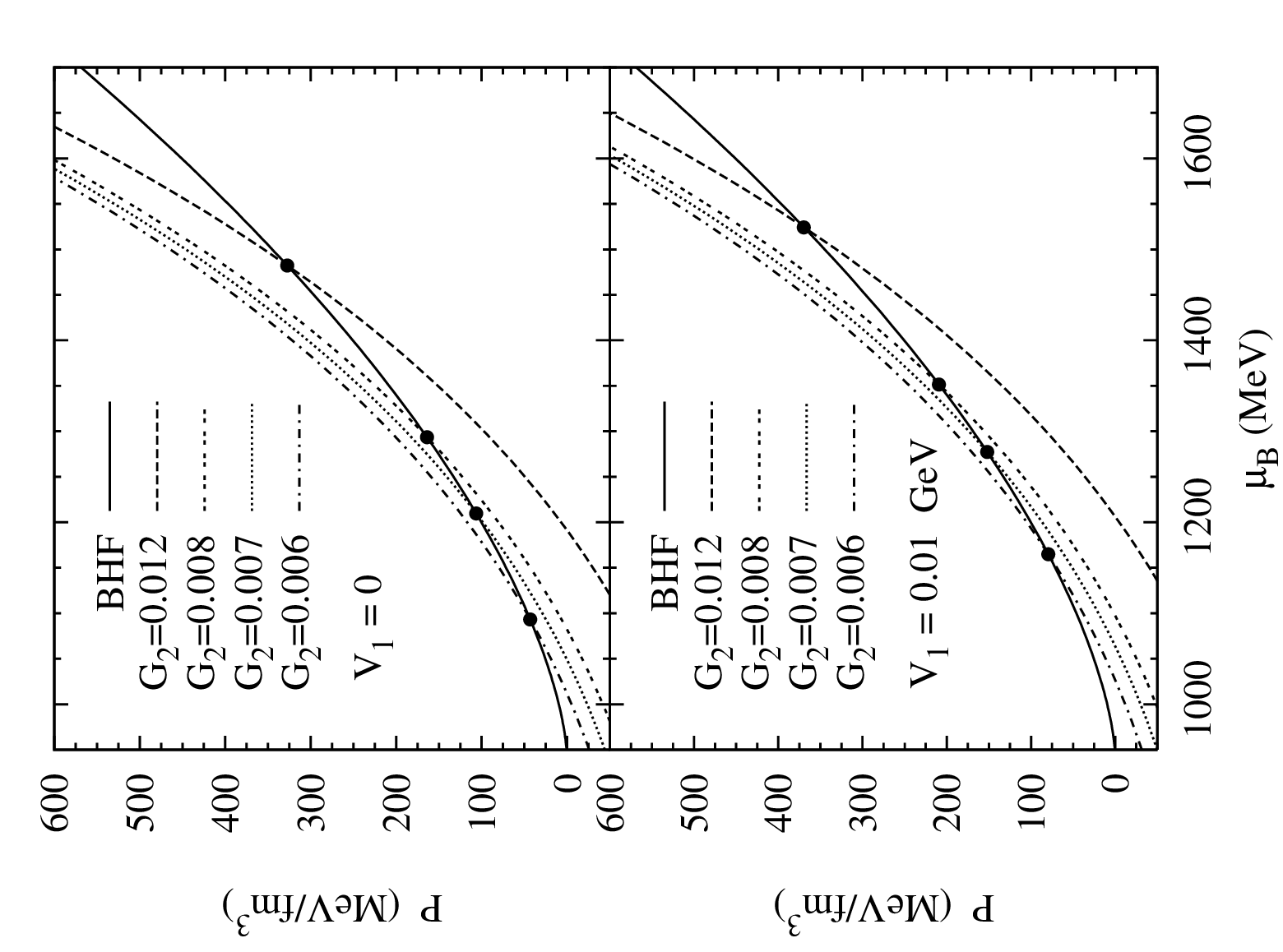}}
\put(200,200){
  \includegraphics[height=5.2cm,angle=270]{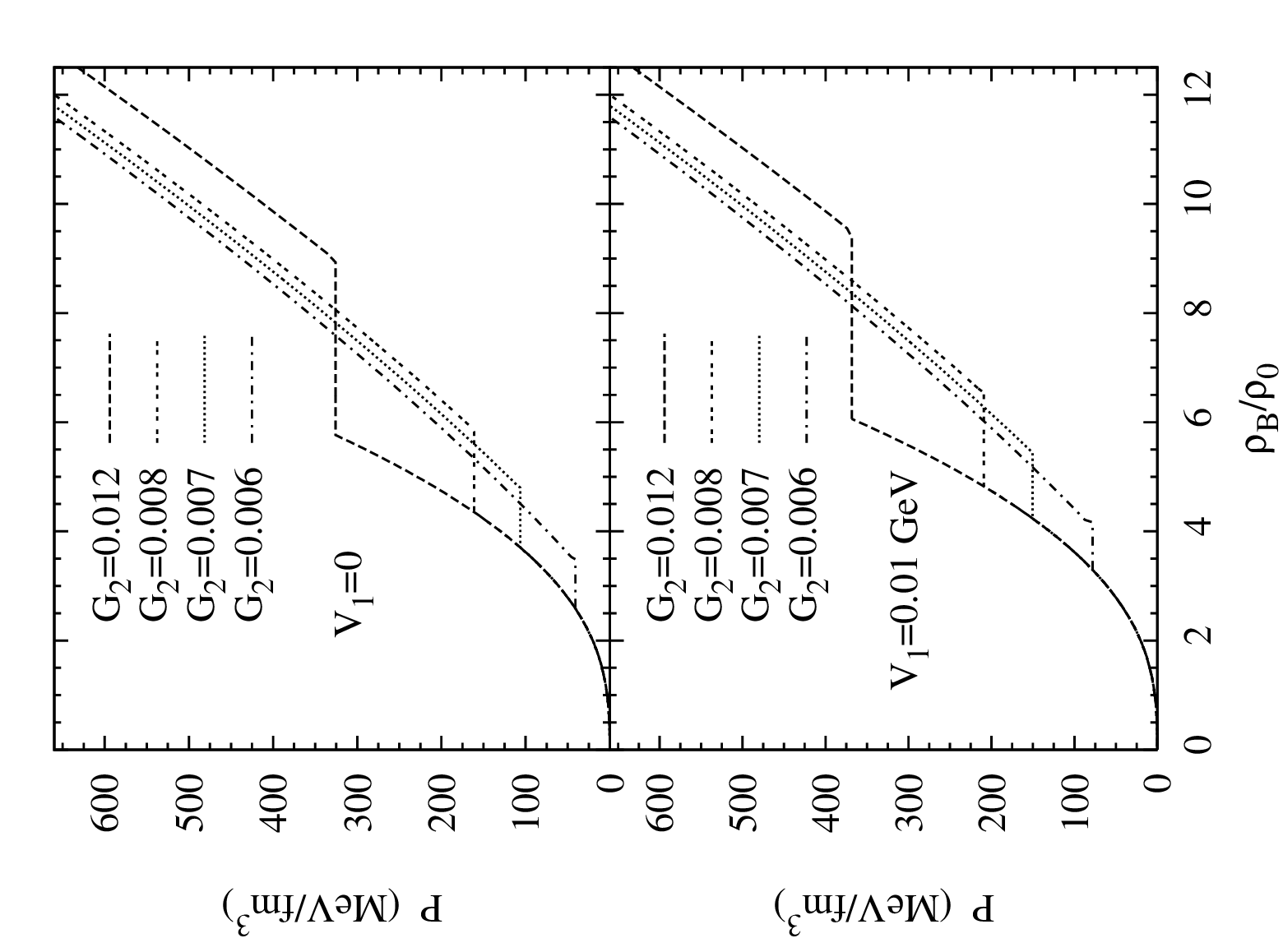}}
\end{picture}
\caption{The pressure vs. baryon chemical is displayed for $V_1=0$ 
(upper left panel), and $V_1=0.01$ (lower left panel). The corresponding
EoS are shown in the right panels.}
\label{f:prho}
\end{figure}

Let us now illustrate the EoS for quark matter. A systematic method to treat 
non perturbative effects in QCD is by gauge invariant field
correlators \cite{phrep}. The approach based on the FCM provides a natural 
treatment of the dynamics of confinement (and of the deconfinement transition) 
in terms of the color electric and color magnetic Gaussian correlators. At $\mu_q=0$, the analytical results are in reasonable 
agreement with lattice data \cite{sim1,sim4,sim22}.
The extension in ref. \cite{sim4} of the FCM to finite values of the chemical 
potential, allows to obtain a simple expression of the Equation of State of 
the quark-gluon matter, which reads
\begin{equation}
\label{pqgp1}
P_{qg} =P_g+\sum_{j=u,d,s} P^j_{q} + \Delta \epsilon_{vac}
\end{equation}
where $P_g$ and $P^j_{q}$ are respectively the gluon and quark pressure,
and
\begin{equation}
\label{pqgp2}
\Delta \epsilon_{vac}
\approx - \frac{(11-\frac{2}{3}N_f)}{32} \frac{G_2}{2}
\end{equation}
corresponds to the difference of the vacuum energy density in the two phases,
being $N_f$ the flavour number, and $G_2$ the gluon condensate whose numerical 
value is known with large uncertainty, 
$G_2=0.012\pm 0.006~ \rm{GeV^4}$ \cite{gluecond}.
Within the FCM, the quark pressure, for a single flavour,
is given by \cite{sim22,sim5,sim6}
\begin{eqnarray}\label{pquark}
P_q/T^4 &=& \frac{1}{\pi^2}[\phi_\nu (\frac{\mu_q - V_1/2}{T}) +
\phi_\nu (-\frac{\mu_q + V_1/2} {T})] \\
\phi_\nu (a) &=& \int_0^\infty du \frac{u^4}{\sqrt{u^2+\nu^2}} \frac{1}{(\exp{[ \sqrt{u^2 +
\nu^2} - a]} + 1)}
\end{eqnarray}
where $\nu=m_q/T$, and
$V_1$ is the large distance static $Q \bar Q$ potential.
The gluon contribution to the pressure is
\begin{equation}\label{pglue}
P_g/T^4 = \frac{8}{3 \pi^2} \int_0^\infty  d\chi \chi^3
\frac{1}{\exp{(\chi + \frac{9 V_1}{8T} )} - 1}
\end{equation}
\noindent
The potential  $V_1$ is assumed to be independent on the
chemical potential, and this is partially supported by lattice simulations
at small chemical potential \cite{sim22}. 
The comparison between the pressure in the two phases is shown in
Fig.~\ref{f:prho} (left panels).
We adopt the simple Maxwell construction, which
implies that the phase coexistence is determined by a crossing point in the
pressure vs. chemical potential plot.
In the upper panel we show the results obtained using $V_1=0$,
whereas in the lower panel calculations with $V_1=0.01$ GeV are displayed.
The solid line represents the calculations performed with the BBG method with
nucleons, and the other lines represent results obtained with different
choices of the gluon condensate $G_2$. The chosen values of
$G_2$ give values of the critical temperature in a range between $160$ and
$190$ MeV \cite{sim22}.
We observe i) the crossing point is significantly affected by the value
of the gluon condensate, and only slightly by the chosen value of the
potential $V_1$, ii) with increasing $G_2$, the onset of the phase
transition is shifted to larger chemical potentials.
In Fig.~\ref{f:prho} (right panels) we display the total EoS
for the several cases shown in the corresponding left
panels. Below the plateau,
$\beta$-stable and charge neutral stellar matter is in the purely
hadronic phase, whereas for density above the ones characterizing the
plateau, the system is in the pure quark phase.

\begin{figure}[t] 
\centering
\includegraphics[width=5cm,angle=270]{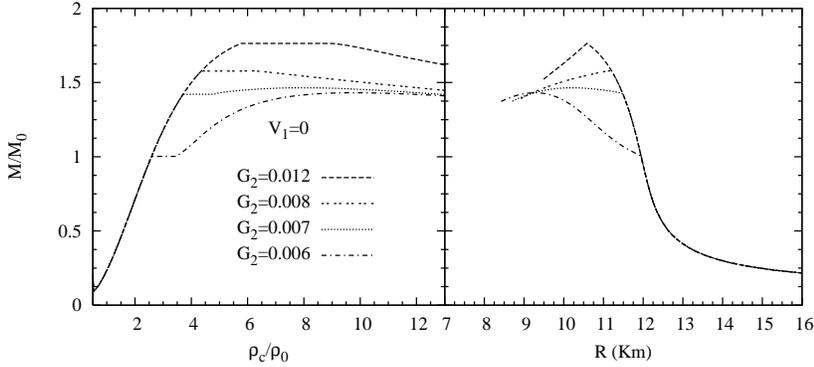}
\caption{
The gravitational mass (in units of the solar mass) is displayed as a function
of the central baryon density, normalized with respect to
the nuclear matter saturation density $\rho_0$ (left panel), and the
corresponding radius (right panel).}
\label{f:mrho_v0}
\end{figure} 

The EoS is the fundamental input for solving the
hydrostatic equilibrium equations of Tolman, Oppenheimer, and Volkov
\cite{shapiro}.
In Fig.\ref{f:mrho_v0} (left panel) we display the gravitational mass
(in units of solar mass $M_\odot = 2\times 10^{33}g$) as a
function of the central baryon density (normalized with respect to the 
saturation value), and the corresponding radius (right panel). We observe that 
the value of the maximum mass spans over a range between 1.4 and 1.8
solar masses, depending on the value of the gluon condensate $G_2$.
The stability of the pure quark phase appears
only for small values of $G_2$, which are hardly in agreement with observational data. In
fact, we recall that any
``good'' equation of state must give for the maximum mass at least 1.44 solar mass, the best
measured value so
far \cite{taylor}. By increasing the value of $G_2$, the maximum mass increases as well, up
to about 1.8 solar
mass, but the stability of the pure quark phase is lost, and the maximum mass 
contains in its interior at most a mixed quark-hadron phase. By switching on 
the potential $V_1$ we observe a similar trend \cite{noi2}. 
However, the observational data indicate that
NS with a mass of at least 1.6 solar masses do exist \cite{vela}, and this puts
a serious constraint on the value of the
gluon condensate, which is not easy to reconcile with the value 0.006 GeV$^4$, 
extracted from the comparison
with the lattice data on the critical temperature. 
As far as the value of the large distance static $Q \bar Q$
potential $V_1$ is concerned, in the comparison with lattice  calculations 
\cite{sim4} one finds a value $V_1\sim 0.5$ GeV at the critical temperature 
and for $\mu = 0$.  We have therefore changed the strength of $V_1$ from  
small values up to $0.5$ GeV, and found that 
already for $V_1 = 100$ MeV the phase transition cannot occur in NS, 
which is then composed of baryon matter only, with a maximum mass around 2 
solar masses. For higher values of $V_1$ the transition can possibly
occur only at exceedingly high values of the density, and therefore the 
quark phase is irrelevant for NS physics.

These results indicate a direct link between the NS quark content and the 
properties of deconfinement in the hadron-quark phase transition. More 
quantitatively, if one considers that the well established values of
NS masses never exceed $\approx 1.6$ solar masses, then these observational 
data constrain  $V_1$ to small
values and in a narrow range, well below 100 MeV, in sharp contrast with 
values around $0.5$ GeV extracted from
lattice calculations. Despite the FCM is in good agreement with full QCD 
lattice data and is a well defined
theoretical approach where confinement is, ab initio, the crucial dynamical 
aspect, some refinements seem to be needed once the astrophysical data are 
considered.


\begin{thebibliography}{99}

\bibitem{bubtes} M. Buballa, Phys. Rep. {\bf 407} (2005) 205.

\bibitem{col} C. Maieron, M. Baldo, G. F. Burgio, and H.-J. Schulze,
Phys. Rev. {\bf D 70}, (2004) 043010.

\bibitem{phrep} A. Di Giacomo, H.G. Dosch, V.I.Shevchenko and Y.A. Simonov,
Phys. Rep {\bf 372}, (2002) 319.

\bibitem{noi1}
M. Baldo, G.F. Burgio, P. Castorina, S. Plumari, and
D. Zappal\`a, Phys. Rev. {\bf C75}, (2007) 035804.

\bibitem{noi2}
M. Baldo, G.F. Burgio, P. Castorina, S. Plumari, and
D. Zappal\`a, Phys. Rev. {\bf D78}, (2008) 063009.

\bibitem{tri} M. Baldo, Nucl. Phys. {\bf A782c}, (2007) 410, and 
references therein.

\bibitem{uix}
% THREE-NUCLEON INTERACTION IN 3-, 4- AND INFINITE-BODY SYSTEMS
 J. Carlson, V. R. Pandharipande, and R. B. Wiringa,
 Nucl. Phys. {\bf A401}, 59 (1983).

\bibitem{bbb}
%MICROSCOPIC NUCLEAR EQUATION OF STATE WITH THREE-BODY FORCES AND
% NEUTRON STAR STRUCTURE
 M. Baldo, I. Bombaci, and G. F. Burgio,
 Astron. Astrophys. {\bf 328}, (1997) 274.

\bibitem{sim1} Yu.A. Simonov, Phys. Lett. {\bf B619}, (2005) 293.

\bibitem{sim4} Yu.A. Simonov, and M.A. Trusov, JETP Lett. {\bf 85} (2007) 598.

\bibitem{sim22} Yu.A. Simonov, and M.A. Trusov, Phys. Lett. {\bf B650} (2007) 
36.
\bibitem{gluecond} 
M.A. Shifman, A.I. Vainshtein, V.I. Zakharov, 
Nucl. Phys. {\bf B147} (1979) 448.

\bibitem{sim5} E.V.Komarov, Yu.A.Simonov, Annals Phys. {\bf 323} (2008) 783.

\bibitem{sim6} E.V. Komarov, and  Yu.A. Simonov,
{\it Theory of Quark-Gluon Plasma and Phase Transition}, arXiv:0801.2251.


\bibitem{shapiro} S.L. Shapiro and S.A. Teukolsky,
 {\it Black Holes, White Dwarfs and Neutron Stars}
(John Wiley and Sons, New York, 1983).

\bibitem{taylor}
% DISCOVERY OF A PULSAR IN A BINARY SYSTEM
 R. A. Hulse and J. H. Taylor,
 Astrophys. J. {\bf 195}, (1975) L51.

\bibitem{vela}
H. Quaintrell et al.,  Astron. Astrophys. {\bf 401}, (2003) 313.

\end{thebibliography}
\end{document}